\documentclass[iop]{emulateapj}

\usepackage{graphicx}
\usepackage{amsmath}
\usepackage[]{natbib}
\usepackage[usenames]{color} 
\usepackage{ulem}
\newcommand{\Msun}{\mbox{$\rm M_{\odot}$}}

\newcommand{\etal}{{et al.~}}

\def\apj{{Astroph. J.}}
\def\apjs{{Astroph. J. Suppl.}}
\def\aj{{Astron. J.}}
\def\mnras{{MNRAS}}

\shorttitle{Radial Migration MDFs}

\shortauthors{Loebman \etal}
\begin{document}

\title{Imprints of radial migration on the Milky Way's metallicity distribution functions}

\author{Sarah R.~Loebman\altaffilmark{1,2}}
\author{Victor P.~Debattista\altaffilmark{3}}
\author{David L.~Nidever\altaffilmark{1,4,5}}
\author{Michael R.~Hayden\altaffilmark{6}}
\author{Jon A.~Holtzman\altaffilmark{7}}
\author{Adam J. Clarke\altaffilmark{3}}
\author{Rok Ro\v{s}kar\altaffilmark{8}}
\author{Monica Valluri\altaffilmark{1}}

\altaffiltext{1}{{Department of Astronomy, University of Michigan,
                 1085 S. University Ave, Ann Arbor, MI 48109-1107, USA};        
		 {\tt sloebman@umich.edu}}
\altaffiltext{2}{Michigan Society of Fellows}
\altaffiltext{3}{Jeremiah Horrocks Institute, University of Central Lancashire,
                 Preston, PR1 2HE, UK; 
                {\tt vpdebattista@uclan.ac.uk}}
\altaffiltext{4}{Steward Observatory, 933 North Cherry Ave, Tucson, AZ 85719}
\altaffiltext{5}{{Large Synoptic Survey Telescope, 950 North Cherry Ave, 
                 Tucson, AZ 85719, USA}; 
                {\tt dnidever@lsst.org} }
\altaffiltext{6}{Laboratoire Lagrange, Universit{\'e} C\^{o}te d'Azur, 
                 Observatoire de la C\^{o}te d'Azur, CNRS, 
                 Blvd de l'Observatoire, 
                 CS 34229, 06304 Nice cedex 4, France; {\tt mhayden@oca.eu}}
\altaffiltext{7}{New Mexico State University, Las Cruces, NM 88003, USA}
\altaffiltext{8}{Research Informatics, Scientific IT Services, 
                ETH Z\"urich, Weinbergstrasse 11, CH-8092 Z\"urich, Switzerland}

\begin{abstract} 
Recent analysis of the SDSS-III/APOGEE Data Release 12 stellar catalogue has revealed that the Milky Way's metallicity distribution function (MDF) changes shape as a function of radius, transitioning from being negatively skewed at small Galactocentric radii to positively skewed at large Galactocentric radii.
Using a high resolution, $N$--body+SPH simulation, we show that the changing skewness arises from radial migration -- metal-rich stars form in the inner disk and subsequently migrate to the metal-poorer outer disk.
These migrated stars represent a large fraction ($ > 50\%$) of the stars in the outer disk; they populate the high metallicity tail of the MDFs and are, in general, more metal-rich than the surrounding outer disk gas.
The simulation also reproduces another surprising APOGEE result: the spatially invariant high-[$\alpha$/Fe] MDFs.
This arises in the simulation from the migration of a population formed within a narrow range of radii ($3.2\pm1.2$ kpc) and time ($8.8\pm0.6$ Gyr ago), rather than from spatially extended star formation in a homogeneous medium at early times.
These results point toward the crucial role radial migration has played in shaping our Milky Way.
\end{abstract}

\section{Introduction}
\label{s:intro.tex}

Observationally, it is challenging to explore the chemical evolution of the Milky Way (MW) for several reasons.
Firstly, stars of a given type span a wide range of distances and are therefore ``smeared-out'' in apparent magnitude space.
Secondly, high-resolution optical spectroscopy is only possible for relatively bright stars.
Finally, the majority of stars in the Galactic disk are obscured by dust at visual wavelengths.
Given these obstacles plus the finite range of Hipparcos parallaxes, most past chemical and/or kinematic studies have focused on the Solar neighborhood \citep[\textit{e.g.},][]{Steinmetz2006, Holmberg2009, Fuhrmann2011, Ramirez2013} or high latitudes \citep[\textit{e.g.},][]{Steinmetz2006, Yanny2009} where the dust extinction is low. 
This approach has allowed for the development of useful insight into our local environment and the stellar halo, but it has given only a partial view of the Galactic disk.

Recently, SDSS-III's Apache Point Observatory Galactic Evolution Experiment \citep[APOGEE,][]{Eisenstein2011} set out to characterize the MW disk using a different approach.
APOGEE is a spectroscopic survey that pushes through the dust ``barrier'' by operating in the near infrared.  
In 2015, APOGEE released its third year catalog \citep[DR12,][]{Alam2015,Holtzman2015} containing 140,000 stars, dominated by red giant branch (RGB) stars that probe to large distances ($\sim$10 kpc).
  
These data have already yielded important insights into the structure and chemistry of the disk including the generation of mean metallicity maps over a large radial range \citep[R $<$ 15 kpc, where R is Galactocentric radius;][]{Hayden2014}, the quantification of the radial metallicity gradient outside R$\sim$5 kpc \citep{Hayden2014}, and the confirmation of a prominent bimodality in [$\alpha$/Fe] for stars with low to intermediate metallicities \citep{Anders2014,Nidever2014}.
Additionally, using the red clump catalog \citep{Bovy2014}, \citet{Nidever2014} found that the chemical abundance pattern of high-[$\alpha$/Fe] stars is remarkably uniform throughout $5<$R$<11$ kpc.
\citet{Nidever2014} interpreted this as indicating that the early evolution of the stellar disk was characterized by a similar star formation history from a well-mixed ISM.
Their simple Galactic chemical evolution model for the high-[$\alpha$/Fe] sequence also implied a short gas consumption timescale (\textit{i.e.}, very active star formation early on).  

More recently, \citet[][hereafter H15]{Hayden2015} used $\sim$70,000 RGB stars (spanning $3<$R$<15$ kpc) from DR12 to analyze the metallicity distribution function (MDF) in the mid-plane and found that the shape of the MDF transitions from being negatively skewed in the inner galaxy to positively skewed in the outer Galaxy, with a nearly symmetric MDF in the Solar neighborhood.
H15 found that heating and epicyclic motion does not alter the shape of the MDF enough to explain the APOGEE observations; rather heating broadens but does not alter the skewness of the MDFs.  
H15 conclude that the positive skewness in the outer Galaxy is most plausibly explained by the movement of more metal-rich stars from the inner Galaxy outward via radial migration.

Radial migration due to transient spirals was first described by \citet{Sellwood2002}. 
In this process, energy and angular momentum changes occur from interactions with transient spiral arms, which move stars at the corotation resonance inward or outward in radius while preserving their nearly circular orbits.
Stars migrated by this process do not experience significant heating because of it \citep{Roskar2013}, while stars migrated by other processes do \citep{Minchev2010}.
Because old stars in the thin disk near the Sun have on average a small epicyclic radius \citep[$\Delta R\sim1.3$ kpc,][]{Sellwood2002}, indicating they haven't experienced significant heating, we investigate the former type of migration rather than the latter.
This has been studied both analytically \citep[\textit{e.g.},][]{Schoenrich2009a, Daniel2015} and in numerical simulations \citep[\textit{e.g.},][]{Roskar2012, Solway2012, VeraCiro2014, Grand2015}.
Migration in the simulation we consider here is dominated by spirals, making it an ideal testbed for comparison to the APOGEE data.
\citet{Roskar2008a} previously demonstrated that this simulation reproduces the MW's MDF in the Solar cylinder; here we extend this analysis to the full APOGEE footprint.

In what follows, we revisit two of the recent APOGEE results: the changing shape of the Galactic MDF in the mid-plane and the constancy of the high-[$\alpha$/Fe] MDF in all spatial volumes.
Using a numerical simulation, we provide strong support for H15's interpretation for the origin of the change in the MDF skewness and provide a novel explanation for the homogeneity of the high-[$\alpha$/Fe] sequence.
That is, we show that \textit{both} results are a natural consequence of a \textit{single} persistent secular process: radial migration, of an early-forming population generated in a limited volume in the case of the [$\alpha$/Fe]-rich population. 
From this we conclude that radial migration likely significantly influenced the Milky Way throughout its history.

\section{Simulation}
\label{s:simulation}

In this paper, we analyze the same simulation that was used in \citet{Loebman2011}.  This simulation is a re-run of the fiducial case discussed in \citet{Roskar2008, Roskar2008a}; it includes metal diffusion but is otherwise identical to the earlier model. 

A detailed description of the simulation's implementation can be found in \citet{Roskar2012}.
Briefly, the simulation is evolved in isolation using the $N$--body+smooth particle hydrodynamics code GASOLINE \citep{Wadsley2004}.
This allows for a high mass and force resolution (baryonic softening of 50 pc).
The initial conditions include two NFW \citep{Navarro1997} halos each containing $10^6$ particles, one of dark matter with a mass of $10^{12}$ $\Msun$ and the other of gas in hydrostatic equilibrium with a mass of $10^{11}$ $\Msun$. 
The gas is set with an initial rotation with $\lambda=0.065$.\footnote{
In earlier works, we reported $\lambda=0.039$.  
Here we have recalculated $\lambda$ for the gas only according to Equation 1 in \cite{Debattista2015}.}
As the simulation progresses, the gas cools and collapses into a disk that forms stars according to a standard prescription: star particles form from gas with a temperature $< 1.5 \times 10^4$ K and a density $> 0.1$ amu/cc \citep{Stinson2006}.

A prescription for stellar feedback is included which injects energy into the surrounding gas and pollutes it with supernova ejecta. 
GASOLINE includes prescriptions for the generation of metals in supernova type Ia and II explosions as well as in asymptotic giant branch stars. 
SN II yields are taken from \citet{Raiteri1996}, SN Ia yields from \citet{Thielemann1986}, and the mass returned to the ISM via stellar winds follows \cite{Weidemann1987} \citep[for details on the metal enrichment implementation see][]{Stinson2006}.
The diffusion of metals through the ISM included a subgrid model of eddy turbulence based on the local smoothing length and velocity gradients \citep{Smagorinsky1963, Wadsley2008}.

As in \citet{Loebman2011}, we define high-[$\alpha$/Fe] stars in the simulation as star particles with [O/Fe] $\ge -0.05$.
This threshold is set lower than the APOGEE threshold ([$\alpha$/Fe] $\ge 0.18$) for two reasons: one, unlike the nascent Milky Way disk, the simulation starts with unenriched gas and two, there is uncertainty in the adopted yields.  
These discrepancies result in a relative offset at the end of the simulation of $\sim0.2$ dex, but as we showed in \citet{Loebman2011} and \citet{Roskar2013} the global trends in enrichment match expectations for the Milky Way.
Therefore, throughout this work, we will focus on the relative trends in the MDFs, which are much more robust, rather than absolute comparison.

At the end of the simulation, the galaxy contains more than $2 \times 10^6$ star particles.
The final disk scale length is $2.5$ kpc, and the disk has a break radius of $10$ kpc.
As done in \citet{Roskar2013} and \citet{Loebman2011}, stars that have moved radially by more than 2 kpc from their birth location are classified as radial migrators, while stars that have moved less than 2 kpc are classified as in situ stars.  
While $25\%$ of the stars in the entire simulation have radially migrated, the relative fraction of migrators varies with radius.
We have verified that the majority of migrators remain on nearly circular orbits, and therefore migrated stars do not reach the Solar cylinder purely because they are heated.
For instance, at the end of the simulation, in the Solar cylinder 80\% of the migrators and 92\% of the in situ star particles have a circularity $x=J_{z}/J_{c}$  greater than $0.90$ (corresponding to an average and maximum epicyclic radius of $1.4\pm0.2$ kpc and $1.9$ kpc respectively), while beyond the disk break (R$>10$ kpc) 75\% of the migrators and 99\% of the in situ star particles have circularity above $0.90$.
 
\section{Results}
\label{s:discussion}

\subsection{MDF of All Stars}

As noted in the previous section, stars in this model undergo significant radial migration; for example, $55\%$ of stars ending in the Solar cylinder ($7.5 \leq $ R$/$kpc $\leq 8.5$) have moved in excess of 2 kpc from their birth radii \citep{Roskar2008a}.  
Thus, the MDF of stars in a particular radial bin will evolve with time due to two effects: ongoing in situ star formation, and the arrival of stars that formed in other regions of the disk with a different chemical evolution history. 

\begin{figure}[!h]
\hskip -.3in
\epsscale{1.2}
  \plotone{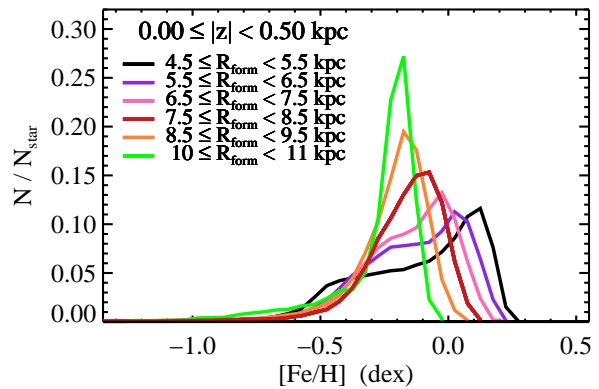}
  \caption{The metallicity distribution function (MDF) as a function of formation radius.}
  \label{f:rform_mdf}
\end{figure}

Figure~\ref{f:rform_mdf} shows the MDF of stars as a function of their radius of \textit{formation}. 
We can think of this as a control case: it is the MDF we would expect if there was no heating or radial migration within the simulation. 
As expected, the MDF of stars forming in the interior of the disk is peaked at higher metallicities than the MDF of stars that form in the exterior of the disk.
Regardless of radial bin considered, the generic shape of the formation MDF is the same: it is peaked at relatively high metallicities and has a long tail toward lower metallicities.  
This shape, with skewness toward the low metallicities, is a generic prediction of simple chemical evolution models (B.~Andrews et al.~2016, in preparation).

H15 found that the skewness of the APOGEE MDFs at large radii is not well described by this distribution.  
Instead, a long tail of high metallicity stars is observed at large radii for stars selected in the plane of the disk (see H15, bottom panel, Figure 5).
H15 showed that if heating effects are included, the MDF is broadened, but the shape of the distribution, including the sense of its skewness, is maintained (see H15, \S{4.2} and Figure 8). 

\begin{figure}[!h]
\hskip -.3in
\epsscale{1.2}
  \plotone{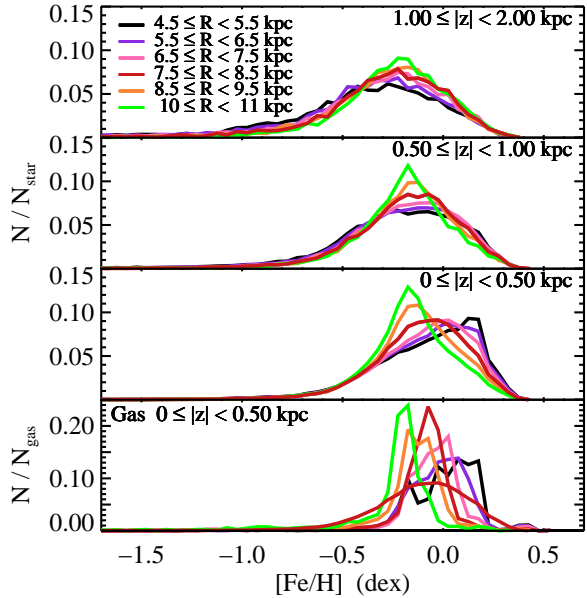}
  \caption{MDF as a function of present day radius for all stars in the simulation.  Varying distances from the mid-plane are considered: (top panel) $1.0 \leq |$z$|/$kpc$ < 2.0$, (top middle panel) $0.5 \leq |$z$|/$kpc$ < 1.0$, and (bottom middle panel) $0 \leq |$z$|/$kpc$ < 0.5$.  The gas MDF is also shown for comparison (bottom panel); the gas MDF is peaked at lower metallicities than the star MDF at the same height.} 
  \label{f:r_mdf}
\end{figure}

In Figure~\ref{f:r_mdf}, we generate MDFs from stars at \textit{the end of the simulation} after significant radial migration has occurred.  
Here we see the same global trends as is seen in the APOGEE data.  
Particularly, in the plane of the disk, the shape of the distribution shifts from being skewed toward low metallicities at small radii, to having no skewness in the Solar cylinder, and finally being skewed toward high metallicities at large radii.
Beyond $\sim$5 kpc, gas in each radial bin (bottom panel of Figure~\ref{f:r_mdf}) is not metal rich enough to have produced the high metallicity stars. 
When volumes away from the plane of the disk are considered, the skewness toward high metallicities decreases in the same fashion as in the APOGEE data (see H15, middle and top panels, Figure 5). 

\begin{figure}[!h]
\epsscale{1.2}
  \plotone{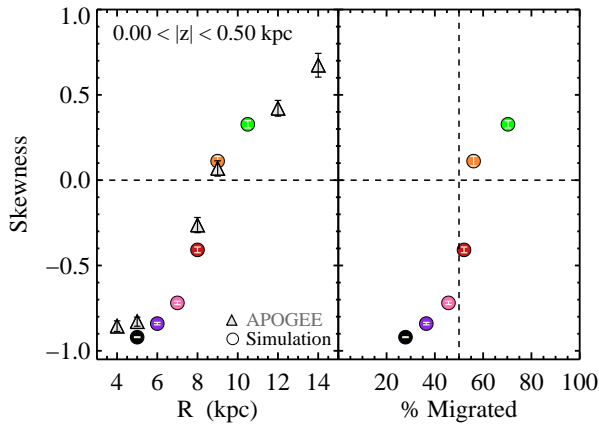}
  \caption{Left: the skewness of simulation (circles) and APOGEE (triangles) MDFs as a function of present day radius within the mid-plane ($0 \leq |$z$|$/kpc$ < 0.5$). Right: the trend in the simulation between skewness and the fraction of migrated stars; note, the color of the circles between the two panels are the same.}
  \label{f:skew}
\end{figure}

For a more systematic comparison between the simulation and observations, we can quantify the skewness of each MDF with a single number.\footnote{
There is a substantial literature on the optimal way to measure skewness \citep[see, for instance,][]{Doane2011}.
So we are not unduly influenced by outliers, we opt to fit a skewed normal model \citep{Azzalini2009} to each MDF, and utilize the skewness parameter that is returned by the best-fit model.
This method is less sensitive to outliers and provides a good fit to the data.
For a sanity check, we have compared the trends in our skewness to those from other non-parametric metrics, and found consistent results.}
Figure~\ref{f:skew} illustrates the skewness measured in each radial bin within the mid-plane ($0 \leq |$z$|/$kpc $ < 0.5$).  
The trend in the simulation shows remarkable agreement with the APOGEE results; in both cases, there is a steady increase from negative skewness to positive skewness with increasing radius with the turnover occurring at the Solar radius.
The right panel of Figure~\ref{f:skew} illustrates the fraction of stars that are migrated in each radial bin in the mid-plane of the simulation.
Note, the larger the radii considered, the larger the fraction of migrated stars, and the larger the positive skewness.
While it is tempting to read off the migrated fraction for APOGEE given a particular skewness, we note that the maximum APOGEE skewness exceeds the maximum simulated skewness.
Thus, only the trend should be considered, rather than the absolute migrated fraction, when drawing comparisons between the APOGEE skewness and the simulation.

\begin{figure}[!h]
\hskip -.3in
\epsscale{1.2}
  \plotone{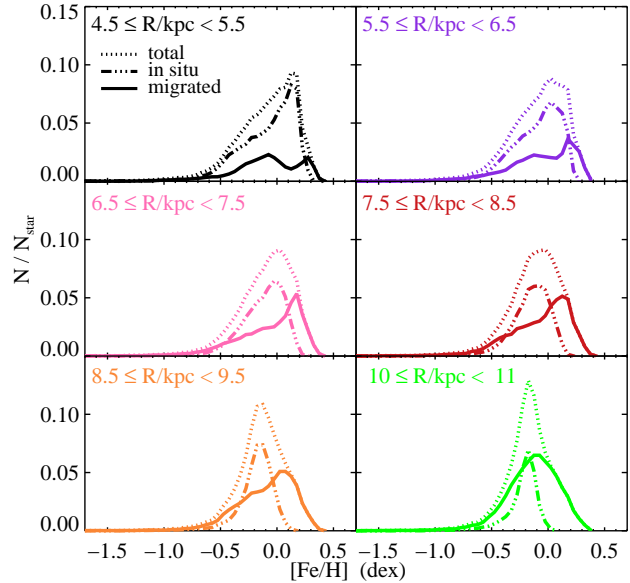}
  \caption{MDF for the total (dotted line), in situ (dashed-dotted line), and migrated (solid line) populations. For ease of comparison, all three MDFs within each panel have been normalized to the total number (in situ + migrated) stars within the volume.}
  \label{f:mig}
\end{figure}

Why is the simulation able to reproduce the observed shift in the shape of the MDF?  
In Figure~\ref{f:mig}, we look in detail at the in situ and migrated stars in the plane of the disk.
At all radial bins, the peak of the total MDF is set by the peak of the in situ population. 
However, beyond $\sim5$ kpc, the peak of the migrated population's MDF occurs at a metallicity where there are very few in situ stars -- in these volumes, the gas is too metal poor to produce them. 
Thus, the stars with the highest metallicities are necessarily migrated, and it is offset between the peak of the in situ and the peak of the migrated population that generates the high metallicity tail.  

We see from Figures~\ref{f:skew} that with increased radius comes an increased fraction of migrators.
As the fraction of migrators rises, the MDF becomes more positively skewed.  
This is a result of the net outward flow of migrated stars.
While it is equally likely for radial migration to move an individual star inward or outward in radius, given the exponential nature of the disk, the net movement of migrators is outward.
Thus MDFs in the outer regions of the galaxy are polluted by metal-rich stars that form in the interior of the disk.
Over time, radial migration builds up the metal-rich tail of the MDF.

\subsection{MDF of High [$\alpha/Fe$] Stars}

Using red clump stars from APOGEE, \citet{Nidever2014} found that the high-[$\alpha$/Fe] chemical abundance patterns (in the [$\alpha$/Fe] vs.~[Fe/H] plane) are the same (within~10\%) throughout the Galaxy.  
\citet{Nidever2014} interpreted the uniformity as an indication that the early Galaxy had a well-mixed interstellar medium (ISM).  
That is, a well-mixed layer of gas was present throughout the Milky Way at early times and experienced a high specific star formation rate throughout forming stars with a similar chemistry everywhere.

\begin{figure}[!h]
\hskip -.3in
\epsscale{1.2}
\plotone{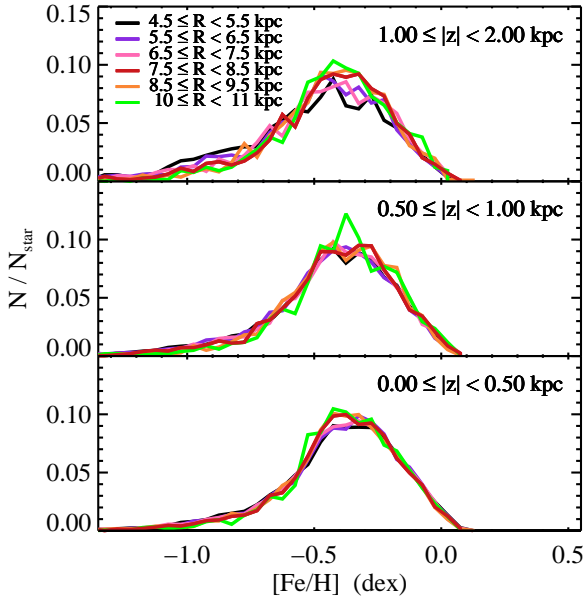}
  \caption{Identical to Figure~\ref{f:r_mdf} for the high-[$\alpha$/Fe] stars.}
  \label{f:r_mdf_alpha}
\end{figure}

H15 revisited the high-[$\alpha$/Fe] APOGEE stars and generated MDFs for this population (see Figure 6 in H15).
Consistent with \citet{Nidever2014}, H15 found remarkable uniformity in the high-[$\alpha$/Fe] MDFs.
In the simulation, we also select high-[$\alpha$/Fe] stars by imposing a flat cut on [O/Fe] (see \S\ref{s:simulation} for details).  
Figure~\ref{f:r_mdf_alpha} shows the resulting MDFs for this high-[$\alpha$/Fe] population.
Without any fine-tuning, the simulation generates the same global fingerprint as seen in H15's high-[$\alpha$/Fe] population. 
Regardless of the radius considered, the high-[$\alpha$/Fe] MDF has a remarkably similar shape, peak, and width; a two-sample Kolmogorov-Smirnov test between all the MDFs in the mid-plane finds that the average probability that these MDFs are drawn from a single distribution is $>90\%$.
Why is this the case?

In the simulation, the high-[$\alpha$/Fe] stars formed in one small region ($3.2\pm1.2$ kpc) in a narrow timespan ($8.8\pm0.6$ Gyr ago) in the inner Galaxy and then migrated out to their present location, all over the Galaxy.  
Because they were formed in a relatively small region of well-mixed gas, these stars are chemically alike; for this reason, regardless of the present day radius considered, the high-[$\alpha$/Fe] MDF looks identical. 
It is worth noting that Figure~\ref{f:r_mdf_alpha} shows a shallow ($-0.03$ dex$/$kpc) negative vertical gradient in the peak metallicity of the high-[$\alpha$/Fe] stars, as stars with $|$z$|\ge 1$ kpc are slightly more metal-poor than stars close to the plane. 
This trend is identical to what is seen in H15; while all the high-[$\alpha$/Fe] stars formed at roughly the same time, the stars at the greatest distance from the mid-plane formed toward the low end of the radial distribution ($\sim3$ kpc) and the early end of this period ($\sim9$ Gyr ago) when relatively less metal-enrichment had occurred.

High-[$\alpha$/Fe] stars are necessarily generated in regions of vigorous star formation.
This environment is naturally produced in the innermost region of the exponential disk as inside-out growth progresses.  
While large gas-rich disks can also generate such vigorous star formation across their full extent, the likelihood that the resulting MDFs are chemically identical is small.
Thus for the Milky Way, it is more plausible that vigorous star formation occurred within a relatively small volume of gas that was chemically well-mixed rather than in a globally well-mixed ISM, and the stars that were generated from this gas were then dispersed throughout the Milky Way via radial migration.

\section{Discussion and Conclusions}
\label{s:conclusion}

We have analyzed the MDFs from a simulation of a disk galaxy that experiences significant radial migration and compared them to recent results from the APOGEE survey.  
From this we draw two main conclusions: 
1) In both the simulation and APOGEE data, the skewness of the MDF shifts as a function of radius, most notably in the mid-plane of the disk.  
In the interior of the disk, the shape of the MDF is skewed toward low metallicities and outside the Solar cylinder the shape of the MDF is skewed toward high metallicities.  
This shift in the skewness is a natural consequence of migration by spiral corotation and requires no fine-tuning to reproduce.
2) Regardless of radial bin considered, the MDFs of the high-[$\alpha$/Fe] stars are identical. 
This is because these stars all formed at roughly the same place at the same early time, and subsequently were spread throughout the disk.    
We suggest that comparing the peak metallicity of gas in a volume to the metallicity and kinematics of stars could be useful in identifying outward migrators if the metallicity of the stars are significantly higher than the peak metallicity of the gas.

\section{Acknowledgments}
\label{s:ack}
We thank Eric Bell for the numerous constructive conversations.
SRL acknowledges support from the Michigan Society of Fellows.  
VPD is supported by STFC Consolidated grant \#~ST/M000877/1.
DLN was supported by a Dean B.~McLaughlin Fellowship at the University of Michigan.
JAH and MRH acknowledge support for this research from the National Science Foundation (AST-1109178).
MV is supported by HST-AR-13890, NASA ATP award NNX15AK79G, NSF Award AST-1515001.
Funding for SDSS-III has been provided by the Alfred P.~Sloan Foundation, the Participating Institutions, the National Science Foundation, and the U.S. Department of Energy Office of Science. 
The SDSS-III website is http://www.sdss3.org/. 
SDSS-III is managed by the Astrophysical Research Consortium for the Participating Institutions of the SDSS-III Collaboration including the University of Arizona, the Brazilian Participation Group, Brookhaven National Laboratory, Carnegie Mellon University, University of Florida, the French Participation Group, the German Participation Group, Harvard University, the Instituto de Astrofisica de Canarias, the Michigan State/Notre Dame/JINA Participation Group, Johns Hopkins University, Lawrence Berkeley National Laboratory, Max Planck Institute for Astrophysics, Max Planck Institute for Extraterrestrial Physics, New Mexico State University, New York University, Ohio State University, Pennsylvania State University, University of Portsmouth, Princeton University, the Spanish Participation Group, University of Tokyo, University of Utah, Vanderbilt University, University of Virginia, University of Washington, and Yale University.

\end{document}